\documentclass[journal,transmag]{IEEEtran}

\usepackage{graphicx}

\usepackage{bm}
\usepackage[utf8]{inputenc}
\usepackage{amsmath}

\graphicspath{{figures/}}
\usepackage{placeins}
\usepackage[draft]{fixme}
\usepackage[switch]{lineno}

\usepackage{cite}
\usepackage{tabularx}
\begin{document}

\title{Personalizing deep learning models for automatic sleep staging}

\author{\IEEEauthorblockN{Kaare Mikkelsen\IEEEauthorrefmark{1,2},
Maarten de Vos\IEEEauthorrefmark{2}}
\IEEEauthorblockA{\IEEEauthorrefmark{1} Department of Engineering, Aarhus University, Aarhus, Denmark}
\IEEEauthorblockA{\IEEEauthorrefmark{2} Institute of Biomedical Engineering, University of Oxford, Oxford, UK}
\thanks{Corresponding author: K. Mikkelsen (email: mikkelsen.kaare@eng.au.dk).}}




\IEEEtitleabstractindextext{%
\begin{abstract}
Despite continued advancement in machine learning algorithms and increasing availability of large data sets, there is still no universally acceptable solution for automatic sleep staging of human sleep recordings. One reason is that a skilled neurophysiologist scoring brain recordings of a sleeping person implicitly adapts his/her staging to the individual characteristics present in the brain recordings.  Trying to incorporate this adaptation step in an automatic scoring algorithm, we introduce in this paper a method for personalizing a general sleep scoring model.  Starting from a general convolutional neural network architecture, we allow the model to learn individual characteristics of the first night of sleep in order to quantify sleep stages of the second night. While the original neural network allows to sleep stage on a public database with a state of the art accuracy, personalizing the model further increases performance (on the order of two percentage points on average, but more for difficult subjects).  This improvement is particularly present in subjects where the original algorithm did not perform well (typically subjects with accuracy less than $80\%$). Looking deeper, we find that optimal classification can be achieved when broad knowledge of sleep staging in general (at least 20 separate nights) is combined with subject-specific knowledge. We hypothesize that this method will be very valuable for improving scoring of lower quality sleep recordings, such as those from wearable devices.

\end{abstract}
 }

\maketitle

\section{Introduction}
Good sleep is fundamental for a healthy life and sleep has been suggested to be important for both diagnosis and treatment of various illnesses \cite{Lamberg2004Promoting,Taheri2006Link,Smaldone2007Sleepless,Stickgold2005Sleepdependent,Carr2018Variability}. Traditional analysis of sleep measurements in a sleep laboratory focuses on differentiating different sleep stages according to the AASM guidelines  \cite{AASM2017scoring}, by which one of 5 labels is assigned to each 30 second epoch (WAKE, REM, non-REM 1, non-REM 2, non-REM 3). Conventionally, this labeling is done manually, by qualified electrophysiologists. 

However, with the increased awareness of sleep for the health of the total population and with continuing rise in the number of personal sleep trackers, including mobile EEG devices that can be used for sleep monitoring \cite{Debener2015Unobtrusive,Looney2016Wearable,Mikkelsen2017Automatic}, there is a renewed interest in developing automated routines as the amount of available sleep data becomes quickly too large to be analyzed through visual screening.

 There is a long history for automated methods for sleep staging \cite{LARSEN1970459,ROBERT1998187}, reaching exciting results using large data sets. Noteworthy examples include \cite{Langkvist2012Sleep} investigating unsupervised approaches, while \cite{Sun2017LargeScale} used an impressively large 2000 subject cohort in their analysis,  using a somewhat simpler approach in combining a 1-layer neural network with a hidden Markow model.
 
 Additionally, recent developments in using deep learning architectures \cite{Cen2017Deep,Biswal2017SLEEPNET,Stephansen2017Use,Tsai2016TakagiSugeno,Zhang2017Sleep} allow skipping the tedious approach of carefully defining characteristic sleep features, and offer a potentially powerful tool for big data sets.





Despite the promise of these machine learning approaches, they have not yet incorporated the idea of exploiting person-specific information in order to improve algorithmic performance, despite the fact that intra-subject variation in sleep features is known to be much smaller than inter-subject variation \cite{Finelli2001Individual,Buckelmuller2006Traitlike}, which is a property that is easily taken into account by human scorers. Such approaches of learning a global method and personalizing the model for each individual has already been successfully investigated for related problems, for instance mood recognition \cite{Zhong2017Cross,Zheng2016Personalizing,Lin2017Improving}), seizure detection \cite{Jiang2017Seizure} and general cross paradigm transfer \cite{Hajinoroozi2017Deep}.  The challenge of sleep tracking naturally involves having access to multiple nights of data from the same individual, and thus it would be a fitting extension to explore the potential of personalizing deep learning models for sleep staging.

In this study we investigate the possibility of using transfer learning \cite{Pratt1993DiscriminabilityBased} for deep convolutional neural networks to transform population models into subject-specific personal models by training a convolutional neural network on data from a large set of subjects and fine-tuning the model for each test subject on data from the first night and evaluating it on a following night. We will first demonstrate that a convolutional network outperforms a traditional feature-based approach  when enough data is available, and subsequently quantify the improvement in performance of fine-tuning the model.
 
\section{Data}

For ease of replication, we will develop all the models on the publicly available ’Sleep EDF’ database \cite{Kemp2002Analysis}, from Physionet \cite{Goldberger2000PhysioBank}. This data set contains 2 nights of sleep EEG from 19 subjects (38 nights in total), together with a single night from an additional subject. As home-tracking devices suitable for collecting large numbers of nights will most likely have access to only a few channels, we select for each subject only the 'Fpz-Cz' derivation from the EEG data and a horizontal EOG derivation.

\section{Methods}

\subsection{Network design}
To evaluate the potential of a convolutional neural network, we started from the network presented in \cite{Tsinalis2016Automatic},  and added some minor tweaks, illustrated in Figure \ref{fig:diag1} and described in the following:

\begin{figure*}[htb]
\includegraphics[width=\textwidth]{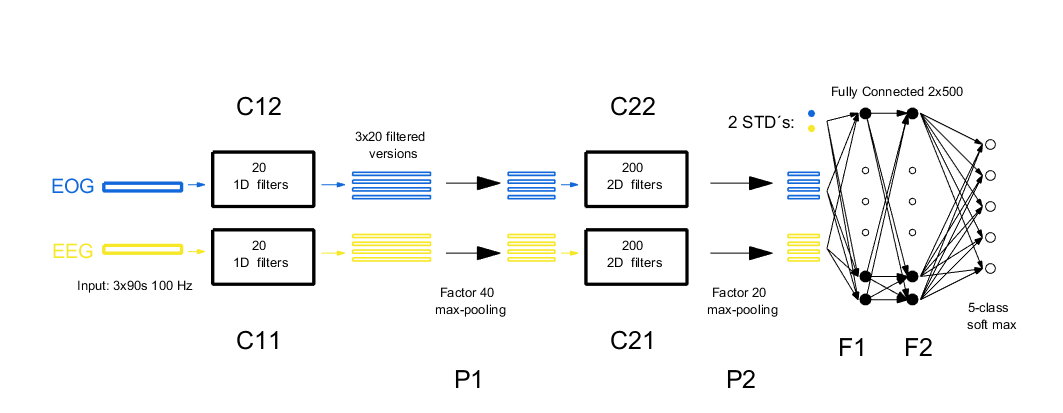}
\caption{Network diagram. Two input streams, EOG and EEG data, are subjected to two different banks of 20 1D filters each. Prior to input to the network, the epoched data is rescaled to always have standard deviation equal to 1. These standard deviations are then passed separately to the system at the fully connected layer, F1.}
\label{fig:diag1}
\end{figure*}

To reduce the memory requirements of the model, some layer sizes were reduced, primarily to have fewer input weights to the fully connected layer 'F1'. Additionally, the network input was expanded in the following manner:

\begin{itemize}
\item An additional input stream was created, with identical setup, analyzing EOG data before passing it to the F1 layer. Since EEG and EOG data are different, the two streams had separate instantiations of the convolutional layers.
\item Instead of passing one epoch of 30 seconds to the network, it now receives 3 epochs, for a total of 90 seconds (for each data stream). This increases performance by giving the network access to the likely states surrounding the epoch in question.
\item Before the data is passed to the network, each epoch is rescaled to have standard deviation (std) equal to 1. The original standard deviations are, however, passed to the network, at layer F1. 
\end{itemize}

Each of these additions provides the network with additional, relevant information about the sleep stage, and was found to increase performance. 

To better facilitate reproduction of our work, a more quantitative description including all parameters is also found in Table \ref{tab:net}. Additionally, the python code will be made available at github (https://github.com/kaare-mikkelsen/sleepFineTuning) upon acceptance of the paper.  

%

\begin{table*}[tp]
\centering
\begin{tabularx}{.7\linewidth}{Xll|X}
Layer (type) & Output Shape & Param $\#$  & Comments\\
\hline 
input1 (InputLayer) & (9000, 1) & 0 &  \\ 
input2 (InputLayer) & (9000, 1) & 0 & \\ 
& & & \\
C11 (Conv1D)   &  (8801, 20) &4020 & \\
C12 (Conv1D)   &  (8801, 20) &4020 & \\
P1 (MaxPooling1D) & (439, 20) & 0& \\
S1 (Reshape) &  (439, 20)&  0&  \\
C21 (Conv2D) &   (410,  200) &  120200 & \\
C22 (Conv2D) &   (410,  200) &  120200 & \\
P2 (MaxPooling2D)& (81, 400) & 0&  \\
inputSTD (InputLayer) &  (2, 1)  &   0 &  \begin{tabular}{@{}c@{}}2 Std's are input \\ alongside output from P2\end{tabular} \\
F1 (Dense) & (500)& 16201500& \\
F2 (Dense)&  (500)&250500& \\
softmax (Dense)& (5) & 2505& 
\end{tabularx} 
\caption{Description of the full network. The total numnber of trainable parameters is 16,578,725. We see that most of the parameters are found in the input to layer F1. During training, 60$\%$ dropout was implemented after layer P2.}
\label{tab:net}
\end{table*}

We implemented the network using Keras \cite{chollet2015keras} and Theano \cite{2016arXiv160502688short} in python 3.5.3, and ran everything on an NVIDIA GeForce GTX 960M with the cuDNN library, version 5005.

When evaluating the performance of the network, we will compare it to a more traditional feature-based approach, more precisely the random forest based method presented in \cite{Mikkelsen2017Automatic}. To do this, we calculate the median accuracy across subjects as a function of amount of subjects (each subject represented by two nights) included in the training set. To keep things simple, we shall exclude the single-night subject from this analysis.
 As a further comparison, we also include single-subject models, in which the classifier is only trained on the first night of each subject, and tested on the second night. For the neural network, the training night is duplicated 20 times (instead of 10, see below), to assure convergence.

\subsection{Network training}

If nothing else is stated, the recording from the subject which only had a single night (as explained in the 'Data' subsection above) was always included in the training data, and never in the test data.


 Each network was trained for 10 'epochs', meaning that all training data was used 10 times, to ensure convergence. 

During training, a 60$\%$ 'dropout' \cite{Srivastava2014Dropout} layer was inserted  between layers 'P2' and 'F1'.

\subsection{Fine tuning}
\begin{figure*}[htb]
\includegraphics[width=\textwidth]{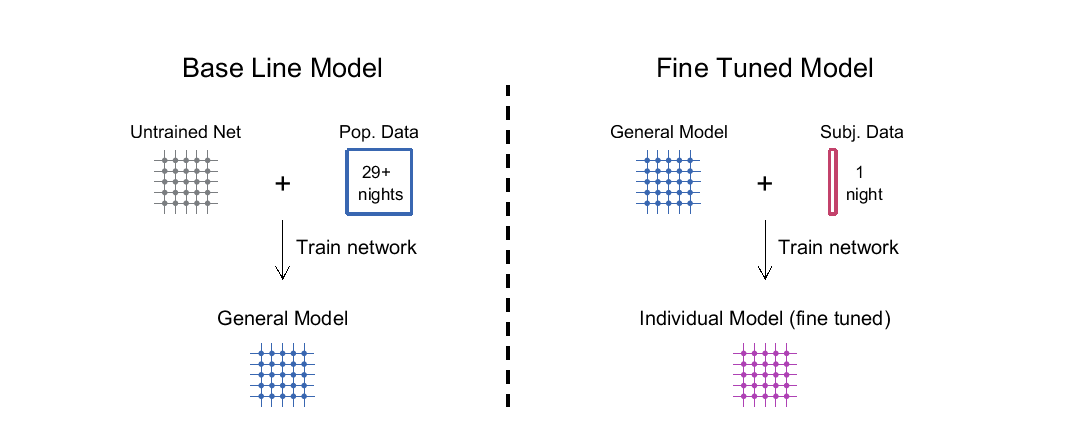}
\caption{Finetuning diagram. First the untrained network is trained using at least 29 nights (depending on the resampling as described above). Afterwards, this general model is subjected to 10 repeated training generations on 1 first-night from a single subject. Finally, the model is tested on the second-night from the same subject.}
\label{fig:diag2}
\end{figure*}

To investigate the improvement that can be obtained by updating the network with one night of subject-specific information, we started from a general population model (a network trained on all subjects), from which a subject specific model was created by 'fine tuning' the general model to the recording from a specific subject. 
 More precisely, in fine tuning, first an untrained network is trained using at least 29 nights (depending on the resampling as described below). Afterwards, this general model was presented with the first-night from a single subject 10 times (meaning this additional training data consisted of 10 duplicates of the one training night). Finally, the model is tested on the second-night from the same subject. See Figure \ref{fig:diag2} for a diagram.
 
As the result of fine tuning might depend on the detailed performance of the base line model, it is important to test the method on a wide range of base line models. To achieve this, we implemented a resampling strategy that exploited maximally the amount of data available as follows:

 During training, all first-nights for all subjects were used, together with $19-n$ second-nights, with $n$ ranging from 1 to 10. This drastically increases the number of possible population models to investigate fine tuning on. As reducing the size of the training set may result in reduced performance of the classifier, we inspected network performance for all iterations and confirmed that the maximal drop in accuracy was acceptably small (average accuracy went from 0.84 to 0.82, when $n$ increased from 1 to 10). Using this resampling scheme, we generated 311 different different baseline models.  The effects of fine tuning was then tested for the different baseline networks and different test data sets, as shown in Figure \ref{fig:scatter_combined}a. This also made it possible to investigate the details of subject specific variations, as shown in Figure \ref{fig:scatter_combined}b and \ref{fig:boxplots}. 
 
In our analysis of the effects of fine tuning, we will test the hypothesis that the stochastic variable defined as the change in accuracy before and after fine tuning has mean value less than or equal to 0, i.e. that fine tuning does not improve accuracy. This will be tested with a standard one-sided t-test.
 

It is worth noting that it was a conscious decision to always include the fine tuning night in the base line training set. This was done to ensure that any resulting improvement in classification accuracy was due to the added knowledge about \textit{which} subject was the most relevant one, rather than the improvement stemming from simply having seen the subject in particular before.

\section{Results}

\subsection{Base line performance}

\begin{figure*}[htb]
\includegraphics[width=\textwidth]{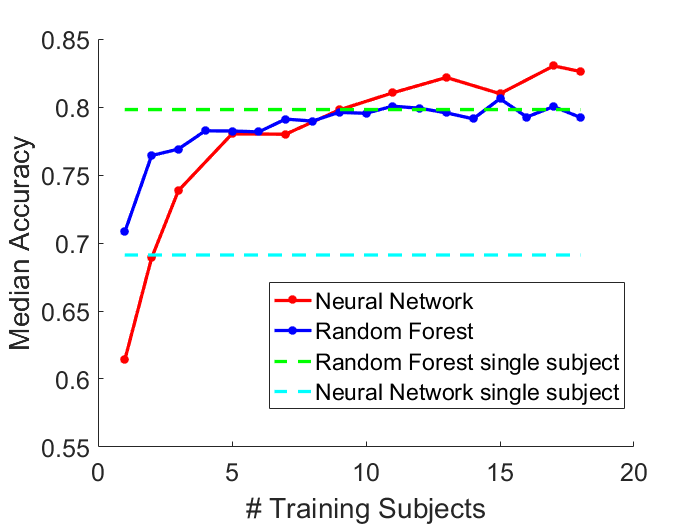}
\caption{Accuracy as a function of number of subjects, for the neural network as well as two different versions of a random forest network. In 'Random Forest single subject' the classifier is only fed one training night and one test night, both from the same subject.}
\label{fig:NN_vs_rf}
\end{figure*}

   Figure \ref{fig:NN_vs_rf} shows the comparison between our neural network classifier and the feature based approach as a function of the amount of training data. We see that when the full data set is used, a very good performance is obtained, and even that the neural network outperforms an individualized random forest classifier. It is also clear that the purely subject-specific neural network classifier performs very poorly. 

\begin{figure*}[htb]
\includegraphics[width=\textwidth]{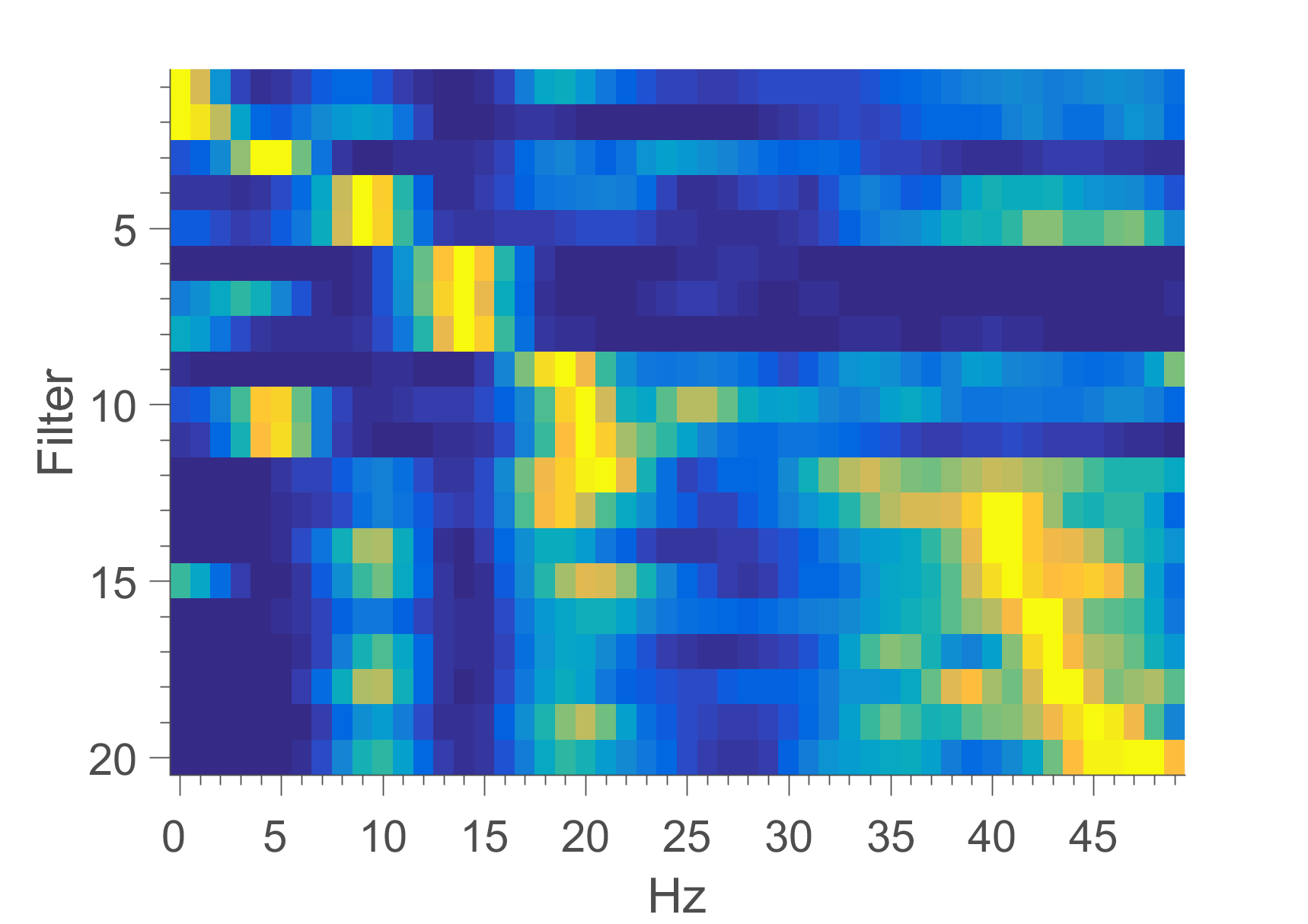}
\caption{Frequency content of 1D EEG filters, generated by calculating the power spectrum of each filter (using the 'pwelch' command in Matlab), and reordering based on the frequency of highest power, to help guide the eye.}
\label{fig:filters}
\end{figure*}

 Figure \ref{fig:filters} shows the frequency content of each of the EEG-based filters. This is calculated by extracting all filter coefficients from layer C11 (the EEG input stream), after training on 36 nights, and estimating the power spectrum of each filter. For the plot, the filters have been reordered according to the frequency of peak power. We clearly see that most of the filters have the form of a band pass filter, and for bands similar to those traditionally used in neuroscience (from \cite{Nunez2007Electroencephalogram}: 'delta': 1-4 Hz, 'theta' 4-8 Hz, 'alpha': 8-13 Hz, 'beta': $>$13 Hz ).


\subsection{Effects of fine tuning}

\begin{figure*}[htb]
\includegraphics[width=\textwidth]{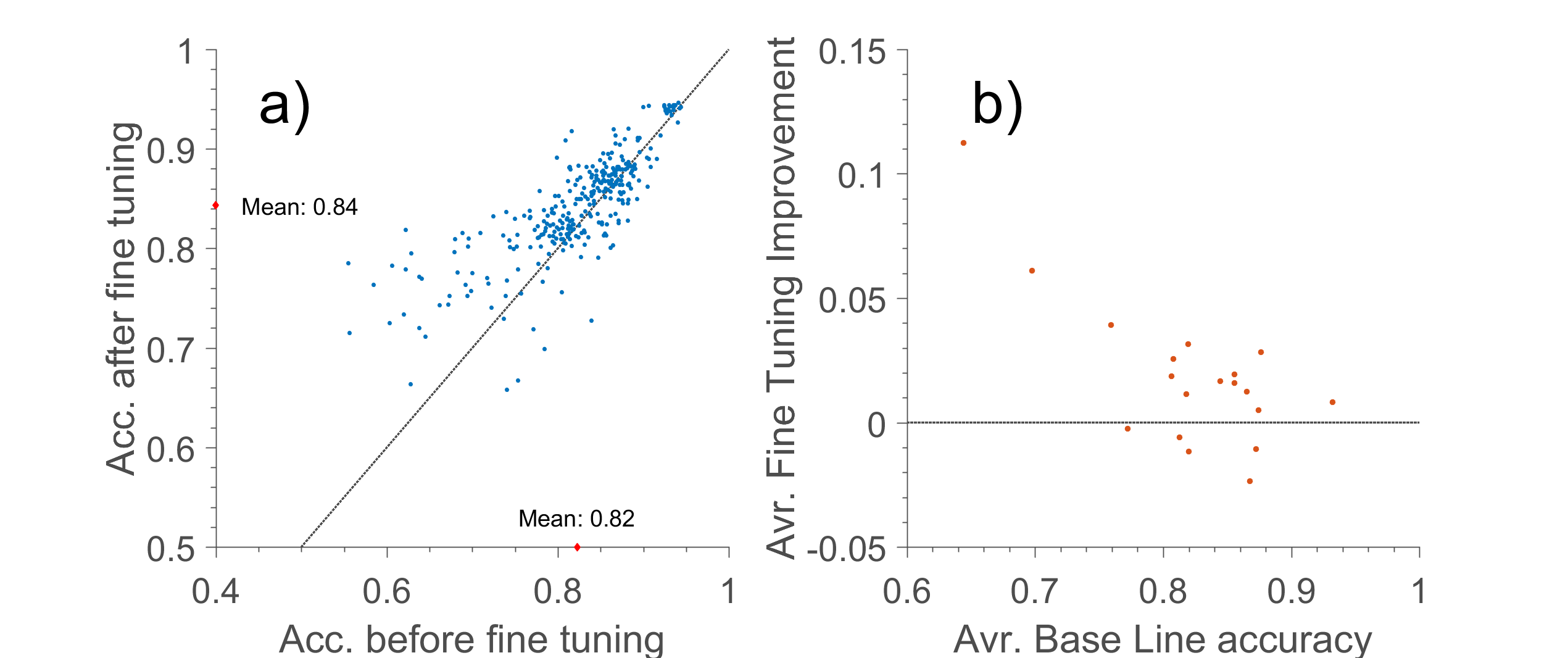}
\caption{A: Scatter plot of accuracy before and after fine tuning. We note that the improvement is particularly large for base line accuracies below 0.8. The line shows x = y diagonal. B: Avr. subject fine tuning. We see that the improvement is much larger for a few subjects.}
\label{fig:scatter_combined}
\end{figure*}

To evaluate in detail the performance improvement relative to the baseline, Figure \ref{fig:scatter_combined} shows two scatter plots. On the left is shown results from 311 different fine tunings, while the right shows average improvement vs. average baseline performance for each of the 19 subjects. In the first case, we see that there is both a general trend towards a slight increase in performance, as well as a more marked effect in cases where the base line performance was relatively poor. In the latter, we see that while an average improvement is present for the great majority of subjects, large improvements primarily happen in a few subjects. Performing t-tests on the results presented in both left and right plots, we find p-values of, respectively, $5\cdot 10^{-15}$ and $0.0076$, meaning that in both cases, the performance increase from fine tuning is statistically significant.

\begin{figure*}[htb]
\includegraphics[width=\textwidth]{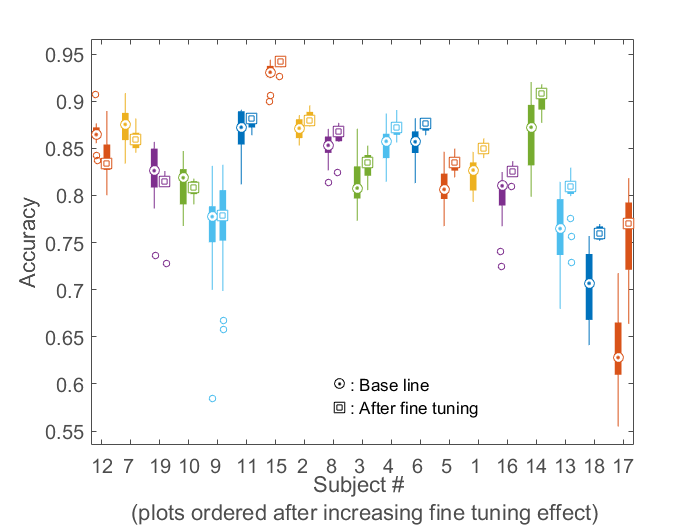}
\caption{Distributions of accuracies before and after fine tuning. Subjects are reordered to achieve increasing fine tuning effect. We note that after fine tuning, the variation in accuracies is generally quite low. Most likely this reflects that fine tuning always uses the same first-night.}
\label{fig:boxplots}
\end{figure*}

 Figure \ref{fig:boxplots} shows the results of fine tuning in a different way: the distributions of baseline and fine tuned performance for each subject is shown. We again note a trend that greater improvement happens when there is more room to improve, and also that the variation after fine tuning is often quite small. This latter fact is likely because we are always fine tuning towards the same first night of the same subject, whereas the resampling of the training data ensures a much greater spread in the base line performance.

\begin{figure*}[htb]
\includegraphics[width=\textwidth]{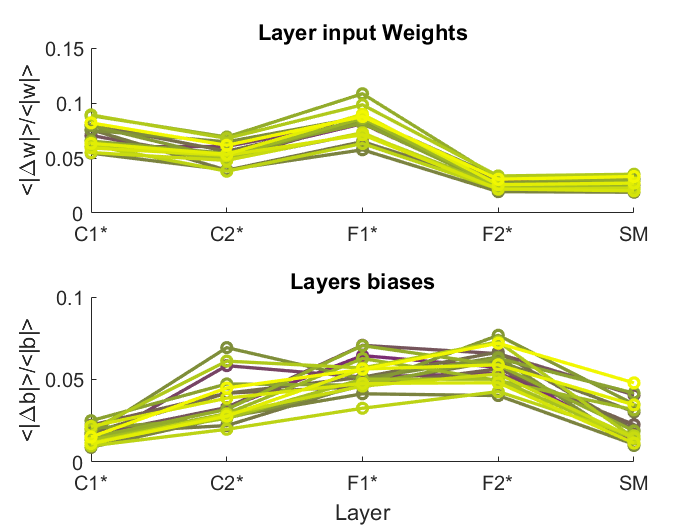}
\caption{Average, relative changes in input weights and biases for each layer. For C11, C12, C21 and C22 is shown the average between the two input streams. Each line corresponds to a subject.}
\label{fig:weights}
\end{figure*}

By comparing network parameters before and after fine tuning, it is possible to study in greater detail where the fine tuning takes place. In Figure \ref{fig:weights} is shown absolute relative change in parameter values before and after fine tuning, averaged across each layer. Each line corresponds to a separate subject. For simplicity, changes in layers C11 and C12 are averaged, as are C21 and C22. We find that fine tuning happens in multiple layers, particularly the input to 'F1' and in 'C11' and 'C12'.

\section{Discussion} 

Starting from a general convolutional neural network architecture, we compare performance to that of a random forest classifier, and demonstrate the network is able to learn the relevant frequency bands for sleep staging, and use these to perform with a high accuracy. For comparison, our network performs at a similar accuracy to other neural network approaches, such as \cite{Tsinalis2016Automatic,Vilamala2017DeepCN}. 

Furthermore, we find that the best sleep staging is achieved when the classifier has access to both large amounts of general sleep data, as well as a weighting towards the individual which the classifier is supposed to analyze data from. We note that the fine tuned network is both better than the individual random forest model showcased in Figure \ref{fig:NN_vs_rf} and both the random forest and neural network general models, despite the fact that these are already quite good. We have also tested the performance of a neural network with only individual data (so one training night and one testing night from the same subject), as seen in Figure \ref{fig:NN_vs_rf} and got results significantly worse than the other alternatives presented here.

Furthermore, we find that the effect of fine tuning on network parameters is spread throughout the network, with a great deal of change taking place in the early layers.
 This is somewhat surprising, given that the conventional wisdom surrounding transfer learning dictates that fine tuning will primarily take place at the last layers. However, in this particular context, it makes good sense; the algorithm is likely looking for the same features (activity in frequency bands), but these are known to exhibit inter-subject variation, and so the early portions of the neural network are tweaked. This is commensurate with the findings of \cite{Hajinoroozi2017Deep}, who also found fine tuning outside of the last layer for EEG data.


In this study, we find that fine tuning the classifier generally improves the performance, especially when base line accuracy is less than 0.8. By delving into the results presented here, we find that much of the baseline variation seen in Fig \ref{fig:boxplots} is related to the inter-subject differences in the data set; if some particular subjects are both included in the $n$ subjects removed for testing, the resulting base line performance is worse, and fine tuning similarly advantageous. Presumably this means that these subjects have some sleep signatures in common, and when that signature becomes less common in the training data, fine tuning increases in efficiency. It is here important to note that the 'rare' sleep features are already in the training set - the fine tuning night does not represent new information, rather it introduces a weighting of the information already available towards those most relevant to the task. On this basis, it is quite possible that fine tuning would retain its effect on larger training sets; knowing the sleep  characteristic of the individual in advance is likely to always be beneficial for the scoring.

 It is beneficial to note that interscorer agreement has been shown to be about $83\%$ \cite{Rosenberg2013American}. This matches our results, particularly those presented in Figure \ref{fig:scatter_combined}, since we are seeing little to no improvement for accuracies above $80\%$. Above this limit, the scoring is likely essentially 'perfect', and the fine tuning does not represent an improvement. Rather, fine tuning fixes those subjects where the classifier is performing much worse than the interscorer agreement. In short, when fine tuning does not work, it is because there is little or no room to improve.
 
Still, it would be interesting to confirm how the results presented here would change for a much larger subject cohort. For comparison, \cite{Sun2017LargeScale} found an increase in performance with increasing cohort size until they reached 300 subjects. From this it would seem that the model presented here, trained on at most 19 subjects, is still highly specialized.

 Finally, it is important to point out that the success of fine tuning is not related to differences between scorers. In the data set used here, 6 different trained scorers were used. In nine subjects, the same person scored both first and second nights. These are subjects 1,5,6,7,10,12,16,18. Comparing this list to Figure \ref{fig:boxplots}, we see that these subjects are not particularly prone to improvement. Indeed, of the 4 subjects in which the median accuracy decreased after fine tuning, 3 of them are on this list.


The data used in this study is PSG data, and as such is likely of a higher quality than what would be available from most mobile sleep monitors - despite the fact that we have here only used two channels. 
We anticipate that in the case of worse starting data (such as mobile EEG), the room for improvement will be larger, and many more subjects may fall in the 'sub $80\%$ category' discussed above. This is similar to the discussion of the 'keyhole hypothesis' put forth in \cite{Mikkelsen2017Keyhole}.

This scenario seems probable no matter the size of the training cohort.

Additionally, we believe that fine tuning would work well for creating classifiers for recordings from people with sleep disorders. In that case, imagine a scenario in which the base line model is trained on a large cohort of healthy subjects, and subsequently fine tuned on a smaller cohort of sick subjects. 

We believe the main drawback of this approach is the need for labeled data for each subject. However, we envisage multiple scenarios in which this is not a serious flaw. Primarily: (a) in many clinical use cases, it would be highly valuable to monitor sleep over an extended period. In these cases, it would not be an issue to have the first night manually scored, even if that should require additional hardware other than the light weight wearable device. (b) it is possible that most of the fine tuning improvement seen here would also be possible using only a scored nap of an hour or so as fine tuning data. In that case, we could imagine a scenario where the patient would come to the clinic to have their wearable handed out, take a day-time nap, and leave with their wearable. In the case where manual scoring requires extra hardware compared to automatic scoring, doing short naps in a controlled setting would likely be a relatively cheap solution.

It is also important to remember that the network architecture used here was not chosen with fine tuning in mind - it seems reasonable to assume that similar gains would be achieved with different architectures.

\section{Conclusion}

 In this paper we investigated the utility of generating personal sleep staging models by fine tuning existing population based models. It was found that this procedure generally increases model performance, particularly for difficult subjects. We theorize that the increase in performance is due to an improved handling of subject-specific quirks.
 
  The main draw back of the tested approach is the need for labeled data from all subjects. However, we believe that there are several, realistic use cases in which this is not a serious issue. 

\section*{Acknowledgments}
The research was supported by Wellcome Trust Centre under Grant 098461/Z/12/Z (Sleep, Circadian Rhythms $\&$ Neuroscience Institute) and the National Institute for Health Research (NIHR) Oxford Biomedical Research Centre (BRC). 

The authors are grateful for the assistance from Orestis Tsinalis in reproducing his results presented in \cite{Tsinalis2016Automatic}.

\section{Bibliography}
\bibliographystyle{IEEEtran}
\bibliography{kbm05}

\end{document}